# Skyrmions in synthetic antiferromagnet nanorings for electrical signal generation

Dimitris Kechrakos[1‡], Mario Carpentieri[2], Anna Giordano[3], Riccardo Tomasello[2†], and Giovanni Finocchio[4*]

[1]*Physics Laboratory, Department of Education, School of Pedagogical and Technological Education (ASPETE), 15122 Athens, Greece*

[2]*Department of Electrical and Information Engineering, Technical University of Bari, 70125 Bari, Italy*

[3]Department of Engineering, University of Messina, I-98166, Messina, Italy.

[4]*Department of Mathematical and Computer Sciences, Physical Sciences and Earth Sciences, University of Messina, I-98166, Messina, Italy*

‡ email: *dkehrakos@aspete.gr*

† email: *riccardo.tomasello@poliba.it*

\* email: *gfinocchio@unime.it*

**Abstract**

Current-driven magnetic skyrmions show promise as carriers of information bits in racetrack magnetic memory applications. Specifically, the utilization of skyrmions in synthetic antiferromagnetic (SAF) systems is highly attractive due to the potential to suppress the Skyrmion Hall effect, which causes a transverse displacement of driven skyrmions relative to the drift direction. In this study, we demonstrate, through analytical calculations and micromagnetic simulations, that in the case of a nanoring geometry, current-driven skyrmions achieve a stable circular motion with a constant frequency, which is a prerequisite for a skyrmion-based clock device. Notably, the operational frequency in a SAF nanoring surpasses that in a bilayer ferromagnetic-heavy metal nanoring and lies in the GHz regime for current densities of 20 MA/cm². We also find that the performance of skyrmions in SAF nanorings is comparable to that of radial Néel domain walls for low current densities (approximately 15 MA/cm²) and low skyrmion densities ($N_{sk}$≈6). Additionally, we introduce a novel skyrmionic three-phase AC alternator based on a SAF nanoring, which operates at frequencies in the GHz regime. Our findings underscore the potential of SAF nanorings as constituent materials in clock devices with tuneable frequencies operating in the GHz regime

## I. INTRODUCTION

The energetic stability and their localized nature of magnetic skyrmions, textures characterized by an integer topological "charge" $Q = (1/4\pi) \int \bm{m} \cdot (\partial_x \bm{m} \times \partial_y \bm{m}) dx dy$ [1, 2], make them a strong candidate as information-bit carriers. [3-12]. In particular, for skyrmion-based racetrack memories a central requirement is the electrical control of skyrmion shifting that has been most efficiently achieved by means of a spin-orbit torque (SOT) [13-19]. The presence of the so-called skyrmion Hall effect (SkHE) [18-20], namely the transverse-to-drift displacement of a current-driven skyrmion, in such devices is an obstacle and different design strategies to suppress this effect have been proposed. Those include engineering anisotropy [21-24], the use of ferrimagnets [25], antiferromagnets [26-29], and synthetic antiferromagnets (SAF) [28,30-35] . The latter consist of two ferromagnetic (FM) layers separated by a non-magnetic (NM) layer that generates an antiferromagnetic RKKY (Ruderman–Kittel–Kasuya–Yosida) coupling between the two FM layers [36]. In SAFs, a pair of skyrmions with opposite topological charges is strongly coupled leading to an overall $Q=0$, and hence to a zero SkHE. Previous studies have also demonstrated that the lack of the SkHE in SAFs is accompanied by very high skyrmion velocities, comparable to those of domain walls (DW) [17, 30, 33].

Although racetrack memory applications usually assume linear racetracks, a nanoring geometry has attracted research interest for either magnetic memory applications [37-40] or for skyrmion-based devices [41] since it has been proved very promising for shift-register applications, such as electric pulse generation based on the periodic circular motion of the skyrmions [42].

Recent experimental studies have demonstrated that an external field gradient in the radial direction could serve as the driving force for skyrmions [44] including ferromagnetic (FM) nanorings [41, 43]. Additionally, micromagnetic simulations have proposed FM nanorings as a versatile platform for various skyrmion-based applications, including a clock with tuneable frequency, a skyrmion-alternator, and a skyrmion-based energy harvester [42]. However, in devices based on skyrmions in FM nanorings, one must account for the Skyrmion Hall effect (SkHE) by appropriately designing the width of the racetrack to stabilize the skyrmion trajectory [41], or by implementing an on-off protocol for the driving current to prevent boundary annihilation of the skyrmions [42]

In this paper, we investigate the dynamics of skyrmions motion in a synthetic antiferromagnetic (SAF) nanoring by using micromagnetic simulations and Thiele's equation. We demonstrate that current-driven skyrmions exhibit very stable circular motion and velocities which are comparable to the ones of DWs. Additionally, we propose a potential application of this scheme for the development of future nano/micro engines based on a skyrmionic three-phase alternator. This would be possible thanks to a proper design of a circulating skyrmion 'train' in a nanoring geometry with three MTJs, necessary for nucleation, shifting and detection of skyrmions, organized in a spatial phase offset of 120 degrees.

## II. DEVICE AND MODELLING

### A. Device Description

The device concept is built on top of a recent experiment demonstrating [33] that skyrmions can be nucleated and manipulated in a synthetic antiferromagnetic (SAF) structure and that the skyrmion detection can be achieved with an MTJ built on top of a skyrmion generator layer. The device here has three main elements: (i) An in-plane pinned ferromagnetic layer (PL) deposited on top of a heavy metal (HM1) with large spin-orbit torque and strong interfacial Dzyaloshinskii-Moriya interactions (IDMI), which stabilizes a radial texture [45] and spin-polarizes radially the vertical electrical current, that is required for the skyrmions to follow a circular motion into the SAF nanoring [5, 30]. We have verified the feasibility of a stable radial magnetization in the PL layer through micromagnetic simulations using the same parameters as in Ref.[45] and the results are shown in Fig.1(c)-(d). The radial magnetization is achieved performing adiabatic switching off of an initially applied saturation field normal to the PL layer. The polar angle of the magnetization ($\varphi_m$) is computed in the mid-width of the nanoring and indicates a continuous transition from a circular arrangement ($\varphi_m = 90^0$) to radial arrangement ($\varphi_m = 0^0$) of the magnetization for a range of IDMI values $1.5 < |D| < 1.9$ mJ/m$^2$. (ii) A SAF skyrmion generation layer [33], composed of two FM layers (FM1, FM2) coupled via a non-magnetic heavy metal layer (HM2) with double functionality, namely generation of antiferromagnetic RKKY interlayer coupling and strong IDMI [36]. (iii) A detection scheme on top of the SAF layer that is based on reading the tunneling magnetoresistive signal generated by a localized magnetic tunnel junction (MTJ1). The MTJ uses the FM2 layer of the SAF nanoring as the free layer [46]. The presence of a skyrmion in the detector region causes a local reduction of the perpendicular magnetization in layer FM2, leading to a substantial variation of the magnetoresistive signal of the MTJ [46-48], which is converted into an output electrical pulse. Two more MTJ detectors are placed on layer FM2 for generation of three-phase ac output signal.

Here, we focus on the current-driven dynamics of Néel skyrmions in the SAF nanoring (Fig. 1), with external diameter ($D$), track width ($w$), and thickness ($t_{FM}$) of each FM layer. Additionally, the localized MTJ detector region of the upper FM layer of the SAF nanoring has an orthogonal shape with width ($w_D$) and length equal to the track width, placed along the radial direction.

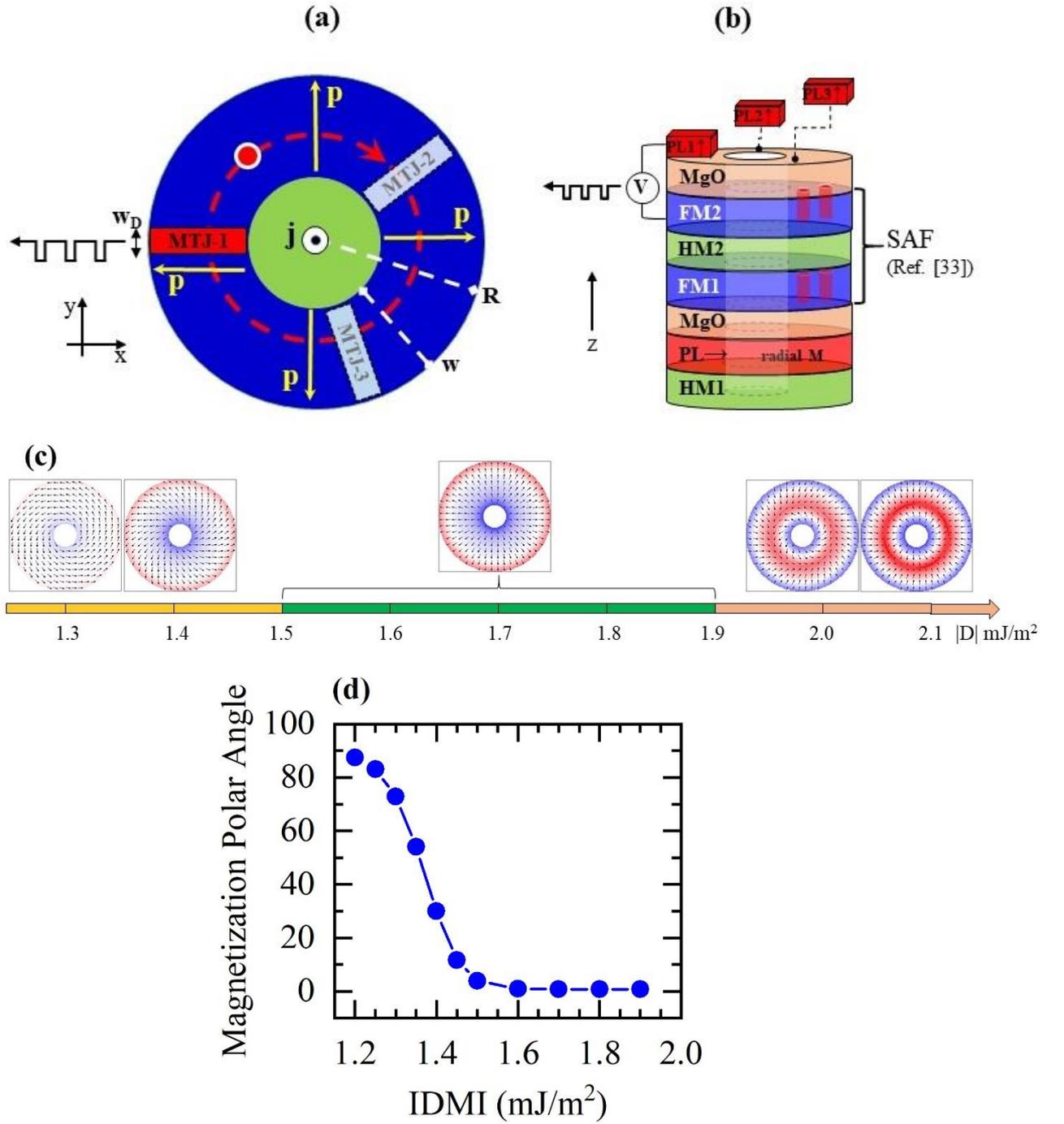

FIG. 1. (a) Top view and (b) 3D view of the proposed skyrmion device for electric pulse generation based on a SAF nanoring. The spin-current along the z-axis is injected into the bottom layer of the device. The current is radially polarized due to the bottom pinned FM layer (PL2) which has strong DMI due to coupling to the bottom HM1. The HM2 layer provides both the required AFM RRKY coupling between the FM1 and FM2 layers of SAF and the required strong interfacial DMI [36]. A localized MTJ on the FM1 layer of the SAF nanoring produces an output electric pulse when a circulating skyrmion crosses the detector region. Two more localized MTJ detectors on the top SAF layer should be placed on FM1 (indicated by dashed lines) to setup the skyrmionic three-phase electrical pulse alternator (see Section IV). (c) Equilibrium configurations of bottom pinned FM layer (PL) at zero temperature and zero applied field indicating the feasibility of stabilizing a radial magnetization configuration for a range of IDMI values ($1.5 < |D| < 1.9$ mJ/m$^2$). Material

4/20

parameters as in Ref [42]. The colors refer to the z-component of the magnetization (blue=-1 and red=+1).(d) Polar angle of in-plane magnetization as a function of the IDMI strength, indicating a continuous transition from circular ($\varphi_m=90^0$) to radial ($\varphi_m=0^0$) spatial distribution of magnetization.

## B. Micromagnetic model

We have used massive micromagnetic simulations performed by means of the state-of-the-art micromagnetic solver *PETASPIN*, which numerically integrates the Landau-Lifshitz-Gilbert equation, augmented by the Slonczewski STT term [49-50]:

$$\frac{d\boldsymbol{m}}{d\tau} = \boldsymbol{m} \times \boldsymbol{h}_{eff} + \alpha_G \left(\boldsymbol{m} \times \frac{d\boldsymbol{m}}{d\tau}\right) - b_J \left[\boldsymbol{m} \times (\boldsymbol{m} \times \widehat{\boldsymbol{p}})\right] \tag{1}$$

where $\boldsymbol{m} = M/M_s$ is the normalized magnetization, $\tau = \gamma_0 M_s t$ is the dimensionless time with the gyromagnetic ratio $\gamma_0 = 2.21 \times 10^5 \, m/A \cdot s$, $\alpha_G$ the Gilbert damping and $\boldsymbol{h}_{eff}$ the dimensionless effective field that includes the external, perpendicular anisotropy, exchange, interfacial Dzyaloshinskii-Moriya (IDMI), magnetostatic and RKKY fields. The RKKY energy density is given as $\varepsilon^{RKKY} = -(A_{ex}/t_{NM})(\boldsymbol{m}^T \cdot \boldsymbol{m}^B)$ and the corresponding interlayer-exchange field is then $\boldsymbol{h}_{RKKY}^{T(B)} = (A_{ex}/t_{NM})\boldsymbol{m}^{B(T)}$, where the indices *T* and *B* refer to the top and bottom FM layers, respectively and $t_{NM}$ is the thickness of the non-magnetic layer. The pre-factor $b_J = \frac{g\mu_B P j}{2\gamma_0 e M_s^2 t_{FM}}$, with $g \approx 2.0$ the Landé factor, $\mu_B$ the Bohr magneton, *e* the electron charge, $t_{FM}$ the FM layer thickness, *P* the polarization coefficient of the injected electrical current and *j* the current density.

The results presented here are for skyrmions in a nanoring with external diameter *D*=252 nm, track width *w*=100 nm, thickness $t_{FM} = 1nm$ and $t_{NM} = 1nm$. Similar qualitative results are also observed for different nanoring sizes. We discretize the ring with cells of dimensions 2×2×1 nm³. We use parameters from our previous works on Néel skyrmions in SAF [10, 30], namely, $M_S$ = 600 kA/m, exchange stiffness *A* =20 pJ/m, interlayer RKKY exchange parameter $A_{ex}$=-0.5 mJ/m², IDMI parameter *D*= 3.0 mJ/m², perpendicular anisotropy constant $K_u$ = 0.6 MJ/m³, polarization coefficient *P*=0.6 and $\alpha_G$ = 0.05. The width of the MTJ detector, if not stated differently, is $w_d = 25nm$, which is approximately equal to the skyrmion diameter. The simulations are performed at zero temperature and without an external magnetic field.

## C. Thiele's Equation

In the case of strong interlayer exchange coupling, the bilayer skyrmion in a SAF propagates in a rigid manner, meaning there is no relative shift between the centers of the constituent skyrmions [51]. The opposite polarities of the two skyrmions cause opposite velocity components along the spin-polarization direction, resulting in the bilayer skyrmion traveling normal to the spin-polarization with complete suppression of the skyrmion Hall effect.

The total topological charge of the bilayer skyrmion is zero due to the opposite skyrmion polarities. We can thus describe the dynamics of the bilayer skyrmion using the well-known Thiele equation [4,52], by neglecting the gyrovector term ($G = 0$) as

$$\alpha_G D \cdot v = F_{STT} , \tag{2}$$

with $v$ the skyrmion (core) velocity, $D$ the dissipative tensor matrix and $F_{STT} = 4\pi B j \, \hat{\varphi}$, the tangential force due to the STT in the current-perpendicular-to-plane (CPP) geometry considered here, exerted by the electrical current with radial spin polarization $\hat{p} = \hat{r}$. The constant $B = (g\mu_B P/2\gamma_0 e M_S^2 t_{FM}) L_{sc}$ with the skyrmion length scale factor $L_{sc} \approx 10^{-9}$ [4]. Solving for the velocity we obtain $v = (4\pi B j/\alpha_G D)\hat{\varphi}$, which describes a uniform circular motion. For a bilayer-skyrmion that circulates at distance $r$ from the nanoring center, the frequency of motion (in Hz) reads

$$f_{SAF} = (\gamma_0 M_s) 2Bj/\alpha_G D, \tag{3}$$

which scales linearly with the injected current density $j$. Finally, comparison to the case of a skyrmion in a FM/HM system [42], assuming the same material parameters and the same trajectory radius, we obtain

$$f_{SAF}/f_{FM} \approx v_\square^{SAF}/v_t^{FM} = (1 + \alpha_G^2)/\alpha_G \gg 1 \tag{4}$$

namely, much higher frequencies are expected in a SAF nanoring compared to its FM counterpart.

### III. NUMERICAL RESULTS

#### A. SAF vs FM nanorings

We initialize the simulation by placing two Néel skyrmions with opposite polarity and the same helicity on the two FM layers of the SAF nanoring with their centers in the mid-width of the nanoring (*r=R-w/2*). We relax the magnetic structure and achieve a stable bilayer-skyrmion. Next we switch on the electrical current for 5ns, which are adequate for the skyrmions to perform at least five laps with the lowest current density used $j = 2 \, MA/cm^2$.

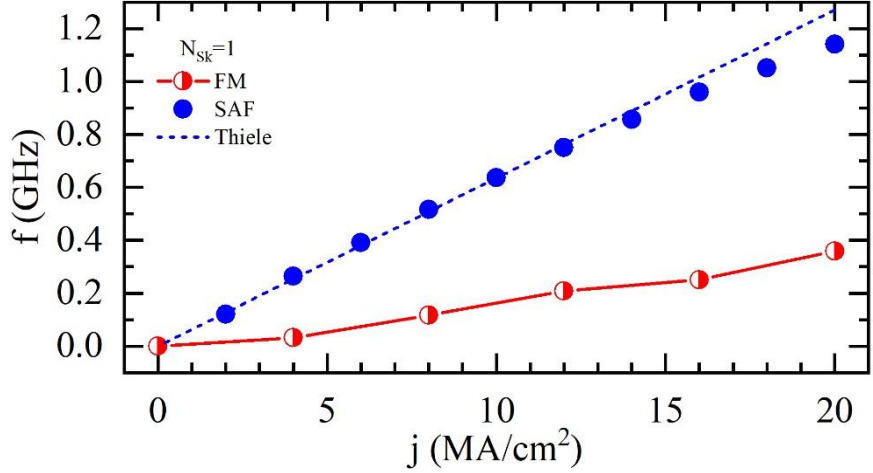

FIG. 2. Dependence of generated frequency on the driving current density. The SAF and FM nanorings have the same size (D=252nm, w=50nm), the same layer thickness t=2nm and the same material parameters (Section IIB). Dashed line shows the analytical results from Eq. (3). For the FM nanoring the electric current has a radial outward direction and scales as 1/r with the distance from the center. The current values for the FM nanoring refer to the inner edge of the nanoring (maximum values).

We compute the frequency of the skyrmion circular motion and to this end, we follow the same procedure as in our previous work [42], namely, we use a post-processing tool to calculate the time evolution of the perpendicular magnetization within the part of the top FM layer that lies in the MTJ detector region (Fig.1a). Enumeration of the dips of the $M_{z,MTJ}(t)$ signal provides the number of laps, $N(t)$, that the skyrmion has performed around the nanoring. Finally, the slope of the $N(t)$ data points is the required circulation frequency of the skyrmion. Obviously, the circulation frequency coincides with the generated electric pulse frequency by the MTJ detector, when a single skyrmion is present in the nanoring. We show in Fig.2 the dependence of the generated pulse frequency on the driving current density for a SAF nanoring. The simulation data are in very good agreement with the predictions of Thiele equation, Eq.(3), except for weak deviations appearing at high currents. These deviations are due to the weak Lorentz contraction [29] of the skyrmions that start appearing for currents above approximately $20 MA/cm^2$. Results for a FM nanoring with the same geometrical and material parameters are also shown for comparison. For those parameters, the SAF nanoring outperforms the FM nanoring by almost 300% for $j = 20 MA/cm^2$ as the generated frequencies of the SAF lie well in the GHz regime. In addition, the skyrmion motion in the FM nanoring is subject to the skyrmion Hall effect, that results in a spiral motion towards the inner edge of the ring and subsequent annihilation as discussed in our previous work [42]. Impementation of a recentering protocol for the skyrmion would cause further reduction of the frequencies in a FM nanoring.

## B. Tuning the frequency in SAF nanorings

We consider here two methods for tuning the generated frequency, namely by varying the number of skyrmions injected in the nanoring and by varying the applied electric current. Consider first the dependence of frequency on the number of skyrmions in the SAF nanoring and the results are shown in Fig.3(a) for fixed current value $j = 20 MA/cm^2$. Skyrmion-chains composed of $N_{Sk} = 1$ up to $N_{Sk} = 10$ are initially injected in the mid-width of the track and let relax without current. The linear scaling of the frequency with number of skyrmions is clearly seen up to approximately $N_{Sk} \approx 7$. As the number of injected skyrmions increases, the skyrmion-skyrmion repulsive forces [5] tend to increase their center-to-center distance and the skyrmions are pushed towards the external border of the nanoring and their size is reduced, as seen in Fig.3(c), due to the skyrmion-boundary repulsive forces. A direct consequence of the reduction of the skyrmion size is the reduction of the exerted STT torque that leads to lower skyrmion velocities. This mechanism explains the sub-linear dependence of the generated frequency on the current density for dense skyrmion chains ($N_{Sk} \geq 7$).

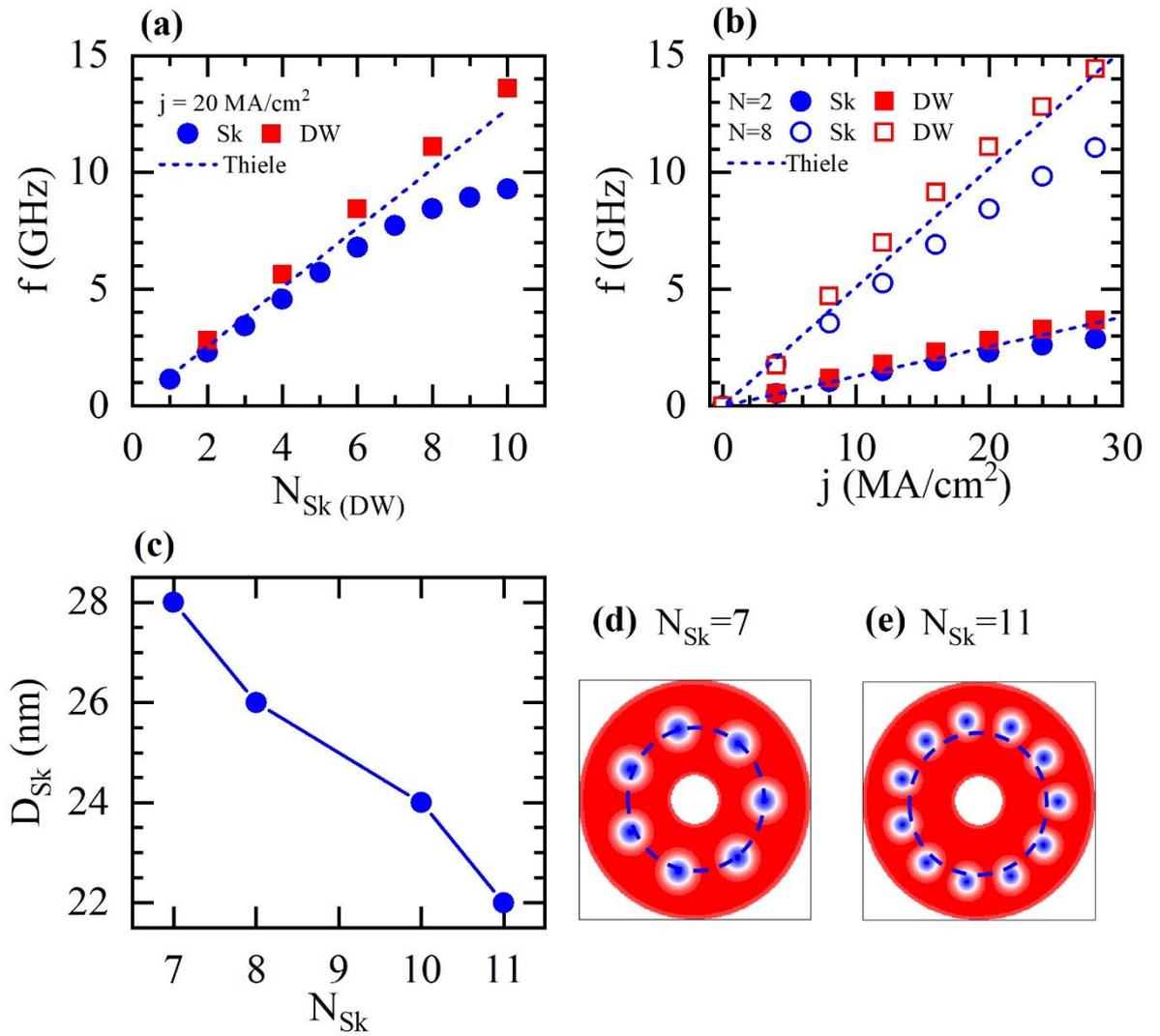

FIG. 3. (a) Dependence of generated pulse frequency on the number of skyrmions and the number of domain walls injected in a SAF nanoring and applied current density $j = 20\, MA/cm^2$. Skyrmions generate approximately the same frequencies as domain walls for low numbers of injected textures ($N_{Sk}$, $N_{DW}$). Skyrmion-skyrmion and skyrmion-edge repulsive forces suppress scalability at large skyrmion numbers ($N_{Sk} \geq 7$). (b) Dependence of generated pulse frequency on the driving electric current density for skyrmions and domain walls. Weakly-interacting skyrmions ($N_{Sk} = 2$) produce similar frequencies to domain walls up to high current values. (c) Dependence of skyrmion diameter on the number of skyrmions circulating in a SAF nanoring. (d-e) Snapshots of nanorings with $Nsk = 7$ and $Nsk = 11$ skyrmion chains. Dashed lines indicate the mid-circle around the track. Size reduction and equilibration towards the outer edge of the ring is seen for dense skyrmion chains.

The generated signal frequency can also be tuned by controlling input current density that drives the skyrmion motion, as Fig.3(b) indicates. When few skyrmions ($N_{Sk} = 2$) circulate in the nanoring, the generate frequenies scale linearly with current density up to high current densities ($j \approx 30\, MA/cm^2$), excpet for weak sub-linearities at high currents due to Lorentz contraction, as discussed above (Fig.2). However in dense chains ($N_{Sk} = 8$) the sub-linear scaling with current is more severe and it can be attributed to a synergy of the Lorentz contraction with skyrmion-skyrmion repulsive forces that introduce a collective response of the skyrmion-chain to the applied current.

## C. Skyrmions versus Domain Walls in SAF nanorings

Néel domain walls along the radial direction can be injected in a SAF nanoring and driven in a rotational motion by electric current. Clearly, in a nanoring only an even numbers of domain walls can be injected, as opposed to skyrmions that even and odd numbers of them can be created.

We compute the signal frequency due to domain walls by counting the number of domain walls crossing the MTJ detector at a certain time interval, in the same way as we do for skyrmions. In Fig.3(a), we show the frequencies for different numbers of domain walls circulating in a SAF nanoring for current density $j = 20 MA/cm^2$. The performance of skyrmions is similar to that of domain walls for small numbers ($N \leq 6$) of textures in the nanoring. For large numbers, domain walls outpace the skyrmions, due to "squeezing" and slow-down of the skyrmions, as discussed above.

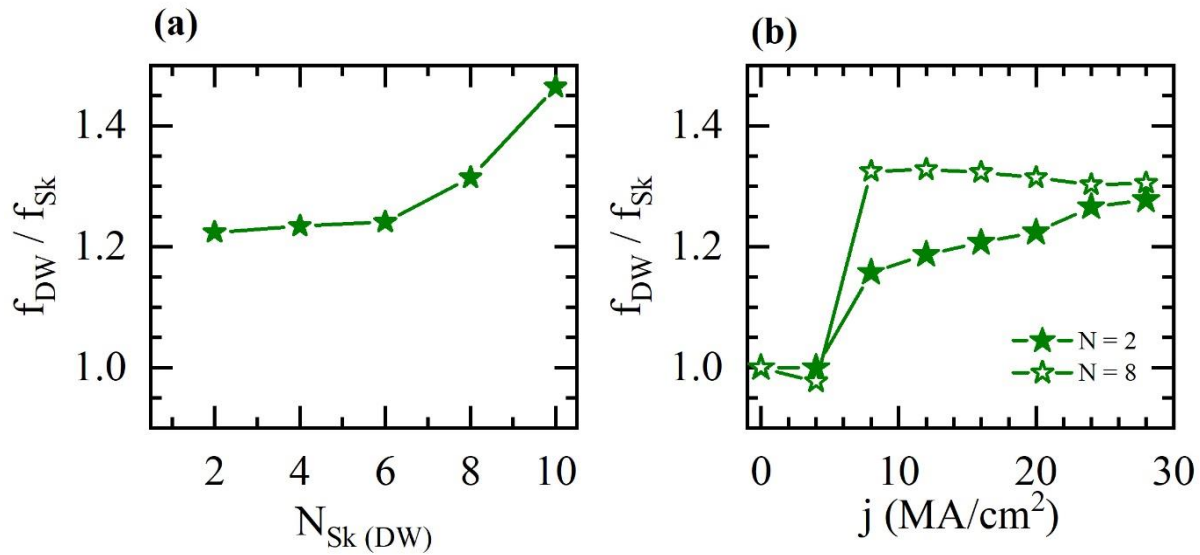

Fig.4. The frequency ratio of circulating domain walls relative to circulating skyrmions in a SAF nanoring as a function of (a) the number of injected textures for current density $j = 20 \, MA/cm^2$ and (b) the driving electric current density. Skyrmions perform similar to domain walls for low numbers ($N_{Sk} \approx 6$) and low currents ($j \approx 10 \, MA/cm^2$).

Furthermore, the frequencies generated when few skyrmions are travelling around the nanoring, as shown in Fig.3(b), remain comparable to those of domain walls for a wide range of applied currents, i.e. up to $30\ MA/cm^2$. However, when the skyrmion chain becomes denser the generated frequencies fall below the domain wall ones with increasing current density. Overall, the performance of skyrmions compared to domain walls in SAF nanorings decreases both with increasing number of skyrmions, Fig.4(a), and increasing current density, Fig.4(b).

## IV.     SKYRMIONIC THREE-PHASE ALTERNATOR

The very stable skyrmion circular motion in SAF nanorings is the key feature for the design of a skyrmionic three-phase alternator. In analogy to a classical balanced three-phase ac alternator [53], the skyrmionic device composed of a SAF nanoring having three MTJs detectors placed at a spatial angle of $120^0$ on top of the last SAF layer (see Fig.1(b)) work as *stator* of the alternator. The chain of skyrmions moving in the SAF nanoring are the *rotor* of the alternator. The skyrmions passing under the three MTJ detectors generate a series of electrical voltages, via the magnetoresistive effect, which constitute the three voltage *phases*.

In Fig.5 we show the dependence of the three-phase signal generated by a SAF with three MTJ detectors (as depicted in Fig.1(b)) on the number of circulating skyrmions. For visual clarity, the three phases have been shifted vertically by their (common) mean value. We observe that with increasing number of skyrmions the peak-to-peak (P2P) amplitude of the signal drops. This happens due to weak reduction of the skyrmion size ("squeezing") with increasing skyrmion density in the nanoring, as discussed previously in Section III.B, in relation to Fig.3. The superposition of the three-phases shows in all cases very low-amplitude oscillations, around a mean value, which shifts gradually from positive values ($N_{sk} = 7$) to negative values ($N_{sk} = 11$). Optimum balance occurs for $N_{sk} = 8$, as the mean value of the oscillations appear close to zero.

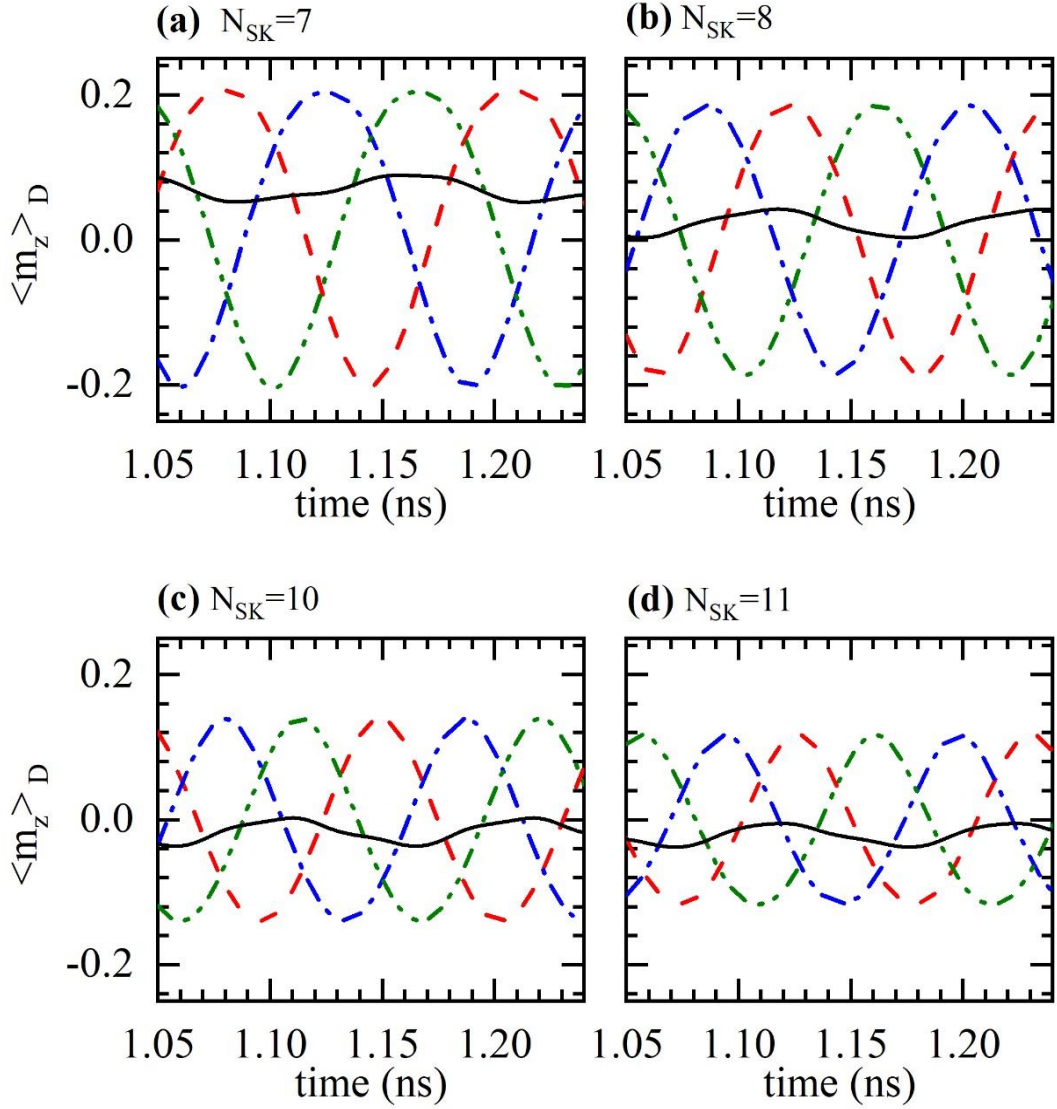

FIG.5. Time-dependence of the spatially average magnetization in the detector region showing periodic pulses generated at three (3) MTJ detectors placed at relative angles of 120° on a SAF nanoring due to circulation of (a) $N_{Sk}$ =7 ($D_{sk}$=28nm), (b) $N_{Sk}$ = 8 ($D_{sk}$=26nm), (c) $N_{Sk}$ =10 ($D_{sk}$=24nm), and (d) $N_{Sk}$ =11 ($D_{sk}$=22nm) skyrmions. All MTJ detectors have width of $w_D$=30nm and the driving current density is j=20 MA/cm². Dash (red), dash-dot (blue) and dash-dot-dot (green) lines are the signals (phases) from the three distinct detectors. The solid (black) line is the superposition of the three signals.

In Fig. 6, we show the dependence of the three-phase signal on the detector width, for $N_{sk} = 7$ skyrmions with (core) diameter $D_{sk} = 28nm$ in the nanoring. Clearly the amplitudes of the three phases drop with the detector width because of the decreasing relative area of the crossing skyrmion to the area of the detector. Also, the mean value of the oscillating total signal, approaches zero (optimum balance) when the detector width is similar to the skyrmion area (Fig.6(c)). As the skyrmion area we understand here the non-saturated area ($m_z < 1$), which is larger than the skyrmion core ($m_z \leq 0$).

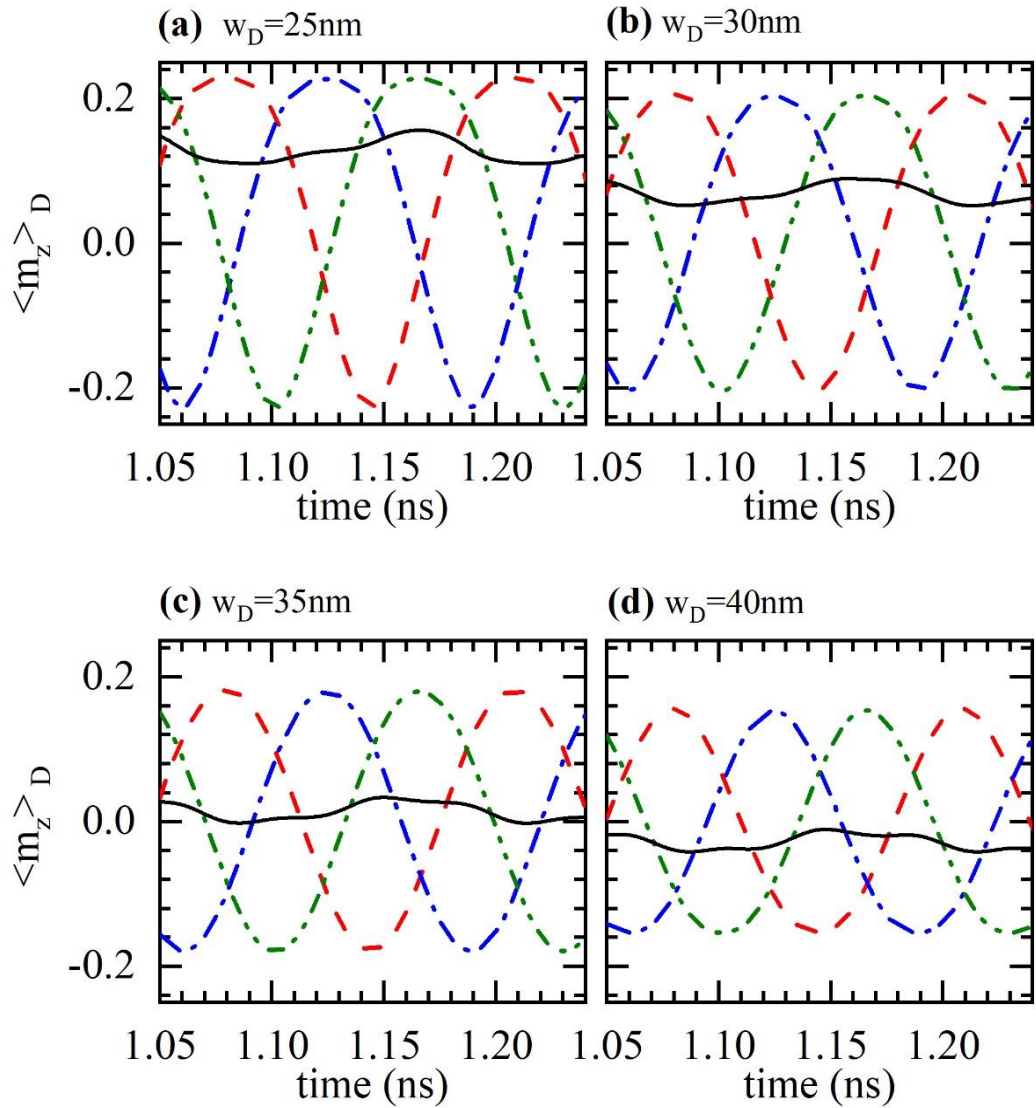

FIG.6. Time-dependence of the spatially average magnetization in the detector region showing periodic pulses generated at three (3) MTJ detectors placed at offset angles of 120° on a SAF nanoring due to circulation of $N_{Sk}$ =7 ($D_{sk}$ =28nm) skyrmions. The detector width is (a) $w_D$=25nm, (b) $w_D$=30nm, (c) $w_D$=35nm, and (d) $w_D$=40nm. The skyrmion (core) diameter is $D_{Sk}$=28nm and the driving current density $j = 20\ MA/cm^2$. Dash (red), dash-dot (blue) and dash-dot-dot (green) lines are the signals from the three distinct detectors ("phases"). Solid (black) line is the superposition of the three signals. The width of the detectors modifies the shape of the three phases and consequently affects the balance of the total signal.

The results shown in Fig. 5 and Fig. 6 demonstrate the successful generation of a three phase ac voltage. However, in a perfectly balanced three-phase ac alternator, the superposition of the three voltage phases is identically zero. In the proposed skyrmionic alternator, we showed that this condition can be fulfilled to a very satisfactory level, by controlling the number of skyrmions ($N_{sk}$) in the chain and the width of the detectors ($w_D$). The center-to-center distance between skyrmions in the chain, essentially modifies the phase difference of the signals from different detectors. For non-vanishing phase difference in the produced signals, the number of injected skyrmions should not be a multiple of $N_{sk} = 3$, as in such a case three skyrmions pass simultaneously through the three detectors and the required phase difference is lost. On the other hand, the width of the detector, modifies the shape of the isolated pulse produced by a single skyrmion. Eventually, the combination of these two factors can be used to optimize the balance of the generated three-phase electrical signal.

Finally, to quantify our previous observations, we define a figure of merit, i.e. the signal quality (SQ), as the ratio of the peak-to-peak (P2P) amplitude of the total signal to the amplitude of the three phases, namely $SQ = 1 - (P2P_{tot}/P2P_{phase})$, which for a perfecty balance three-phase ac signal is equal to unity. We show in Fig.7 the behavior of the signal quality, where it appears that for the SAF nanoring with diameter $D = 252nm$ and track width $w = 52nm$ the optimum choice is a chain of $N_{sk} = 8$ skyrmions and MTJ detector width $w_D = 30nm$.

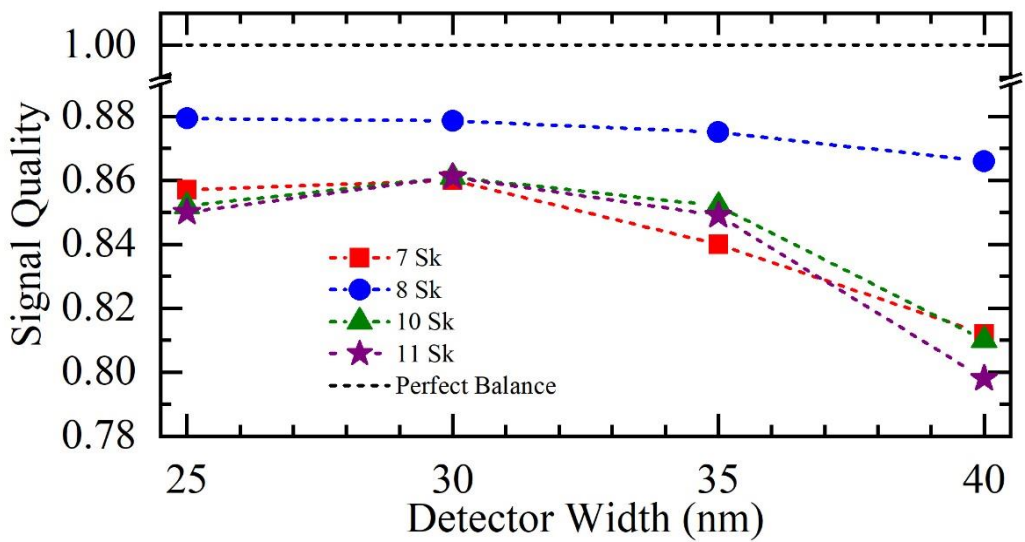

FIG. 7 Dependence of a 3-phase ac signal quality, generated by skyrmion-trains on a SAF nanoring, on the number of skyrmions and the width of the 3 MTJ detectors used. Signal quality is controlled by the number of injected skyrmions in the SAF nanoring and the width of the grown-on-top MTJ detectors.

## V. CONCLUSIONS

We have demonstrated the feasibility of a skyrmionic clock device using a SAF bilayer biased by a dc electrical current that generates electrical pulses in the GHz regime with tunable output signal frequency. Compared to a FM nanoring device, the SAF nanoring device has two main advantages, namely the stability of the generated electric pulse frequency and the superior frequency values. We performed a proof-of-principle of a skyrmionic balanced three-phase current alternator based on a SAF nanoring with three MTJ detectors. For the proposed device, we showed that the balance of the generated signal can be optimized by proper choice of two design parameters, namely the number of skyrmions injected in the nanoring and the width of the MTJ detectors. These results could be very promising for the emergent field of nanomachines, to design more compact power suppliers fostering the miniaturization of nano-robots. Our micromagnetic simulation results point to further applications of SAF nanoscale clocks in the highly-demanded GHz frequency regime.

**Acknowledgements**

DK acknowledges the hospitality by the Politecnico di Bari – Dipartimento di Ingegneria Elettrica e dell'Informazione (DEI) during the course of the work and the financial support from the Special Account for Research of the School of Pedagogical and Technological Education through programs "*Educational and Research Infrastructure Support*" (No 52922) and *"Strengthening of Research in* ASPETE". Useful discussions with Dr. Alejandro Riveros (Universidad Central de Chile) are gratefully acknowledged. The work of RT, MC and GF was supported by the projects PRIN 2020LWPKH7 "The Italian factory of micromagnetic modelling and spintronics", PRIN20222N9A73 "SKYrmion-based magnetic tunnel junction to design a temperature SENSor—SkySens", funded by the Italian Ministry of Research, and by the project number 101070287—SWAN-on-chip—HORIZON-CL4- 2021-DIGITAL EMERGING-01. RT, MC and GF are with the PETASPIN team and thank the support of the PETASPIN association (www. petaspin.com).